\documentclass[12pt]{iopart}
\usepackage{iopams} 
\usepackage{bm}

\usepackage{graphicx}
\newcommand{\beq}{\begin{equation}}
\newcommand{\beqn}{\begin{eqnarray}}
\newcommand{\eeq}{\end{equation}}
\newcommand{\eeqn}{\end{eqnarray}}

\def\be{\begin{equation}}
\def\ee{\end{equation}}
\def\bea{\begin{eqnarray}}
\def\eea{\end{eqnarray}}
\begin{document}

\title{Non-Gaussianity of the density distribution
in accelerating universes II:N-body simulations}

\author{Takayuki Tatekawa$^{\dag,\ddag}$ and Shuntaro Mizuno$^{\|}$}

\address{\dag The center for Continuing Professional Development,
 Kogakuin University, 1-24-2 Nishi-shinjuku,
 Shinjuku, Tokyo 163-8677, JAPAN}
\address{\ddag Advanced Research Institute for Science and Engineering,
Waseda University, 3-4-1 Okubo, Shinjuku,
Tokyo 169-8555, JAPAN}
\address{$\|$ Research Center for the Early Universe (RESCEU),
 School of Science, University of Tokyo,
 7-3-1 Hongo, Bunkyo, Tokyo 113-0033, JAPAN}

\begin{abstract}
We explore the possibility of putting constraints 
on dark energy models with statistical property of
large scale structure in the non-linear region.
In particular, we investigate the $w$ dependence
of non-Gaussianity of the smoothed density distribution
generated by the nonlinear dynamics. In order to follow
the non-linear evolution of the density fluctuations,
we apply N-body simulations based on $P^3 M$ scheme.
We show that the relative difference between non-Gaussianity
of $w=-0.8$ model and that of $w=-1.0$ model  is
$0.67 \%$ (skewness) and $1.2 \%$ (kurtosis)
for $R=8h^{-1}$ Mpc. We also calculate the correspondent 
quantities for $R=2h^{-1}$ Mpc,
$3.0 \%$ (skewness) and $4.5 \%$ (kurtosis),
and the difference turn out to be  greater, even though
non-linearity in this scale is so strong that 
the complex physical processes about 
galaxy formation affect the galaxy distribution.
From this, we can expect that the difference 
can be tested by all sky galaxy surveys
with the help of mock catalogs created by selection functions, 
which suggests that non-Gaussianity of the density 
distribution potentially plays an important role for
extracting information on dark energy.

\end{abstract}

\maketitle

\section{Introduction}\label{sec:intro}

The formation of large scale structure in the Universe
is a central element of astrophysics.
The key idea in understanding the formation of the structure
in the Universe is gravitational instability 
\cite{Peebles80,Peacock,Liddle,Coles}.
Roughly stated,
if the material in the Universe distributed
irregularly, then the overdense regions provide extra
gravitational attraction, draw material toward them
and become more overdense.
At present, this simple picture dramatically succeeds
in explaining what is observed and current attention
is focused entirely on the details of the 
gravitational instability process which strongly
depends on the cosmological models.

One of the useful means to extract the detail 
of structure formation is 
the statistical property of observable quantities.
Among various statistical quantities, 
the density probability distribution
function (PDF) of the density contrast 
plays a very important role because it
includes information of non-Gaussianity.
In the standard picture, 
the primordial density fluctuations originated from quantum
fluctuations are stretched to large comoving
scales during the inflation phase and  assumed to be
random Gaussian \cite{Starobinsky:1982ee}. 
It is, however, well known that
even though the density fluctuations
remain Gaussian in the linear regime, 
it significantly deviates 
from Gaussian once the non-linear stage is reached
\cite{Kofman94,Yokoyama96}.
Since structure formation is
competition between the cosmic expansion and gravity,
the growth of the density contrast as well as 
the evolution of its statistical
quantities are intimately tied to the 
cosmic expansion law of the Universe.

On the other hand,
according to recent observations for type Ia
Supernovae~\cite{Perlmutter99},
the expansion of the Universe is accelerating.
Combining measurements of Cosmic Microwave Background
Radiation~\cite{WMAP03} and recent galaxy redshift survey
~\cite{Tegmark04}, 
we are forced to recognize the existence of 
a cosmological constant, or kind of dark energy
whose value is almost the same order of magnitude as
the present density of the Universe~\cite{WMAP}.
(For reviews about the cosmological constant problem,
see \cite{Weinberg89}.)
From the phenomenological viewpoint, 
in order to clarify the property of dark energy,
it is important to constrain the effective equation of state
of dark energy $w$ by observations. Even though we
have obtained the constraint as $w < -0.90$ (95 \% confidence limit
assuming $w \ge -1$)~\cite{WMAP} by combining WMAP data
with other astronomical data, in order to pin down 
the value of $w$,
it is necessary to propose other independent
and complementary methods.

For this purpose, as stated at the beginning,
we suppose the statistical property of 
large scale structure of the Universe
is helpful. Actually, we were the first to analyze
the $w-$dependence of non-Gaussianity of the PDF
of the smoothed density contrast
in \cite{Tatekawa06} where we considered 
simple constant-$w$ dark energy models 
with $w=-0.5, -0.8, -1.0, -1.2$. 
(The case for the cosmological constant ($w=-1$) had already
been considered in such as \cite{Plionis95,Borgani95,Kayo01}.)
There, in order to see the effects of 
the non-linear dynamics analytically, as a first
step, we adopted Lagrangian linear perturbation
\cite{Zeldovich70,Arnold82,Tatekawa05}  which
 describes the evolution of the density fluctuation
in the quasi-nonlinear region better than 
second order Eulerian perturbation 
\cite{Munshi:1994zb,Yoshisato:1997eb}.

As a natural extension to the previous one,
in order to go into the strongly nonlinear region where 
the difference of non-Gaussianity depending on $w$
is expected to be large, numerical simulations seem
to be necessary. Even though the simulations including
dark energy are important, because of computational expense
of the simulations, the first simulations including dark
energy other than the cosmological constant appeared only
in 2003. By now the aim of N-body simulations
in the presence of dark energy are limited to
the mass function of dark matter halo
\cite{Linder:2003dr,Klypin:2003ug,Kuhlen:2004rw}
as well as the matter power spectrum
\cite{McDonald:2005gz}.
Therefore, in this paper, we analyze the $w$-dependence
of non-Gaussianity
of the PDF of the smoothed density contrast
with N-body simulations
to cover the strongly nonlinear region where the results of
 \cite{Tatekawa06} remain invalid.

This paper is organized as follows. In section
\ref{sec:dark_energy}, we briefly summarize the effects
of dark energy on the structure formation scenario and
define the quantities which are necessary to estimate
non-Gaussianity of the density distribution
in section \ref{sec:non-Gaussianity}.
The set up and results of N-body simulations
are described in Section \ref{sec:N-body}
and Section \ref{sec:results}, respectively.
Section \ref{sec:summary} is devoted to conclusions.

\section{Effects of dark energy}
\label{sec:dark_energy}

Here, we briefly summarize the effects of dark energy.  
We consider almost the same situation
as our previous paper \cite{Tatekawa06} where
we assumed the following facts;

(a) As for matter components, we consider (cold dark) matter
and dark energy, i.e. 
$\rho_{\rm tot} = \rho_m + \rho_{\rm DE}$,
where $\rho_{\rm tot}$, $\rho_m$ and $\rho_{\rm DE}$
are total energy density of matter, each component
of energy density of matter and dark energy, respectively.

(b) Dark energy interacts with matter only gravitationally,
i.e. the energy conservation of matter and dark energy
holds independently.

(c) The effective equation of state of dark energy
$w$ is constant for simplicity, even though 
in principle, we can relax this condition.

(d) Curvature of the Universe is negligible,
i.e.  ($\mathcal{K}=0$).

Then, the background Friedmann equation can be written as
\begin{equation}
H^2 = H_0^2 \left [ \Omega_{m0} a^{-3} + \Omega_{DE0}
 a^{-3(w+1)} \right ] \,, \label{eqn:Friedmann-Omega}
\end{equation}
where $a$ is a scale factor of the Universe,
$H = \dot{a}/a$ is its Hubble parameter.
We have also defined the density parameters for 
matter and dark energy:
\begin{equation}
\Omega_{(m,\;\;DE)} \equiv \frac{8 \pi G}{3 H^2} \rho_{(m,\;\;DE)}
\label{eqn:def-Omega} \,.
\end{equation}
The subscripts $0$ denote that 
the corresponding quantities are 
evaluated at the present time.
From Eq.~(\ref{eqn:Friedmann-Omega}), the effects of  
dark energy affect the background 
cosmic expansion, which depends on the effective equation
of state of dark energy $w$.

Since structure formation is described as competition
between the cosmic expansion and gravity, 
the change of the cosmic 
expansion modifies the growth of the fluctuations.
In the following, we summarize the results obtained in 
structure formation, limiting to the linear perturbation
theory, for simplicity.

In the comoving coordinates, the Poisson equation 
is given by
\begin{equation}
\nabla^2 \Phi = \frac{3}{2} \Omega_{m} H^2 a^2  \delta_m, 
\end{equation}
where $\Phi$ is the gravitational potential in the comoving
coordinates, $\delta_m \equiv (\rho_m - \rho_{mb})/\rho_{mb}$
is the density fluctuation of matter, 
$\rho_{mb}$ is the background density of matter.

Combined with the continuity equation and Euler equation,
the growth equation of the matter perturbation
can be obtained
\begin{equation}
\ddot{\delta}_m + 2H \dot{\delta}_m -
\frac{3}{2}H^2 \Omega_m \delta_m=0, 
\label{growth}
\end{equation}
where the dot denotes a time derivative.

Therefore the growth of large scale structure
depends on the background cosmic expansion law
and it provides one of the constraints to
the property of dark energy.

Strictly speaking, in dark energy models with
a dynamical field, one might wonder whether
perturbations of the field might affect
the growth of structure. However, since calculations
of the galaxy power spectrum in such models suggest
that the field is essentially smooth
on these scales \cite{Coble:1996te,Caldwell:1997ii}, 
we assume that
the perturbations of the field do not impact structure
on these scales.

\section{Non-Gaussianity of the density distribution}
\label{sec:non-Gaussianity}

In what follows, since we are usually interested in 
the density fluctuations of matter, we will 
omit the subscript $m$ unless otherwise stated.
In order to analyze the statistical properties
of the density contrast, we introduce 
a one-point probability distribution function
of the density fluctuation field $P(\delta)$
(PDF of the density fluctuation) which denotes 
the probability of obtaining the value $\delta$.
If $\delta$ is a random Gaussian field, 
the PDF of the density 
fluctuation is determined completely as
\begin{eqnarray}
P(\delta) = \frac{1}{(2\pi \sigma)^{1/2}}
 e^{-\delta^2/2\sigma^2}
\label{Gauss},
\end{eqnarray}
where 
$\sigma^2 \equiv \left < \left (\delta-\left <\delta 
\right> \right )^2 \right > $ 
is the dispersion and $\left <\;\;\right >$ denotes 
the spacial average.

At initial, $\delta$ is often treated as a 
random Gaussian field as a result of
the generic prediction of
inflationary scenario \cite{PNG}.
In the linear perturbation theory, 
Gaussianity of $\delta$
is completely conserved if we start 
with a Gaussian initial condition, 
because each Fourier mode
evolves independently according to the growth law
Eq.~(\ref{growth}). 

Once nonlinear terms are considered,
however, the PDF deviates from the initial Gaussian shape.
The point is that $\delta$ is constrained to have
a value $\delta \geq -1$, otherwise the energy density
$\rho$ would be negative \cite{Coles}.
The Gaussian distribution (\ref{Gauss}) always assigns 
non-zero probability to the unphysical 
regions with $\delta < -1$.
The error is negligible when $\sigma$ 
is small because the probability  
of obtaining the value $\delta <-1$
is then very small, but, as fluctuations enter the strongly
nonlinear regime with $\sigma \sim 1$, this error must
become so important that the Gaussian distribution is only
a poor approximation to the real distribution.
What happens is that, as fluctuations evolve, 
mode-coupling effects cause 
the initial distribution to skew, generating a long 
tail at high $\delta$ while they are bounded 
at $\delta=-1$.

If the PDF deviates from Gaussian distribution, 
the cumulants of the one-point PDF of the density fluctuation
field whose orders are higher than two become nonzero.
Especially, the third and fourth order cumulants,
which are defined as 
$<\delta^3>_c \equiv <\delta^3>$,
$<\delta^4>_c \equiv <\delta^4>-3\sigma^4$
mean the display asymmetry and non-Gaussian degree of
``peakiness'', respectively, for a given dispersion.

Since it is known that the scaling 
$<\delta^n>_c \propto \sigma^{2n-2}$ holds
for weakly non-linear region, if we consider 
the gravitational clustering from Gaussian initial
conditions \cite{Bernardeau02}, we introduce 
the following higher-order statistical
quantities~\cite{Peebles80,Peacock}:
\begin{eqnarray*}
\mbox{skewness} &:& \gamma = \frac{ \left < \delta ^3 \right >_c }
{\sigma^4} \,, \\
\mbox{kurtosis} &:& \eta = 
\frac{\left < \delta^4 \right >_c}{\sigma^6} \,.
\end{eqnarray*}
The merit of adopting these definitions is, as stated above,
that they are constants in extremely weakly nonlinear stage
which are given by Eulerian linear and second-order
perturbation theory \cite{Peebles80,Bernardeau02}. 
For example, 
in Einstein-de Sitter Universe, the skewness
and kurtosis are given by
\begin{eqnarray*}
\gamma &=& \frac{34}{7} + {\cal{O}}(\sigma^2) \, ,
\nonumber\\
\eta &=& \frac{60712}{1323} + {\cal{O}}(\sigma^2) \, .
\end{eqnarray*}

Even though it is shown that the skewness
in weakly nonlinear region does not deviate much
from $34/7$ in several dark energy models 
\cite{Kamionkowski:1998fv,Gaztanaga:2000vw,Benabed:2001dm},
the region we can observe is nearer and the nonlinearity
becomes stronger. Therefore, in the following,
we concentrate on these quantities by analyzing 
the nonlinear evolution of the fluctuations.

It is worth noting that in the previous paper 
\cite{Tatekawa06},
we used other definitions for
the skewness and kurtosis which are based on 
\cite{Kofman94} in which
$<\delta^n>_c$ is divided by $\sigma^n$. 
The ones in the previous definition 
should be divided by $\sigma$ and $\sigma^2$
for the skewness and kurtosis, respectively, to obtain
the ones in the present definition.

\section{N-body simulations}
\label{sec:N-body}

In our previous paper \cite{Tatekawa06},
in order to analyze the non-linear dynamics, we applied
Lagrangian perturbation theory where we can
extract the quasi-nonlinear nature even if we consider
only the linear order. Since we could obtain
the naive  $w$-dependence of non-Gaussianity of 
the density distribution in the weakly nonlinear region,
as a natural extension, in this paper, 
we apply cosmological N-body simulations
in order to obtain quantitatively plausible predictions.

Below we describe the N-body codes, initial conditions,
parameter choices and technical details of simulations.
The results themselves are reported 
in Section~\ref{sec:results}.

The numerical algorithm is applied by particle-particle 
particle-mesh ($P^3M$) method
whose code had been originally written by
Bertschinger ~\cite{P3M}. The initial Gaussian fluctuation 
is generated 
by COSMICS~\cite{COSMICS}.
Until the quasi-nonlinear regime, COSMICS evolves
the fluctuation within the scheme of 
Lagrangian linear perturbation.
Then, we start N-body simulations at $z \simeq 30$.

For N-body simulations, we set the parameters as follows:
\begin{eqnarray*}
\mbox{Number of particles} &:& N=128^3 \,, \\
\mbox{Box size} &:& L=128 h^{-1} \mbox{Mpc}
 ~~(\mbox{at}~a=1)  \,, \\
\mbox{Softening length} &:& \varepsilon = 50 h^{-1} \mbox{kpc}
 ~~(\mbox{at}~a=1)  \,.
\end{eqnarray*}

The cosmological parameters
at the present time ($a=1$) are given
by WMAP~\cite{WMAP}:
\begin{eqnarray}
\Omega_m &=& 0.28 \,, \\
\Omega_{DE} &=& 0.72 \,, \\
H_0 &=& 73~ \mbox{[km/s/Mpc]} \,, \\
\sigma_8 &=& 0.74 \,.
\end{eqnarray}

For dark energy models,
we set several equations of state.
\begin{equation}
p_{\rm DE}= w \rho_{\rm DE}, ~~ (w=-0.5, -0.8, -0.9, -1.0, -1.2) \,,
\end{equation}
where $w=-1.0$ corresponds to the cosmological constant. Even though
$w=-0.5$ has already been excluded and the strong
energy condition is violated in the case of $w=-1.2$,
we calculate these cases as well for comparison.

According to the recent clustering results of XMM-Newton soft
X-ray sources, the parameter $\sigma_8$ depends on $w$
\cite{Basilakos06}.
The relation between $\sigma_8$ and $w$ can be fit by
\begin{eqnarray}
\sigma_8 &=& 0.34 (\pm 0.01) \Omega_m^{-\gamma(\Omega_m, w)} \,, 
 \label{eqn:sigma8-w} \\
\gamma(\Omega_m, w) &=& 0.22 (\pm 0.04) - 0.40 (\pm 0.05) w
 - 0.052 (\pm 0.040) \Omega_m \,.
\end{eqnarray}
Using this relation, we compute $\sigma_8$ for several $w$ cases
(Table~\ref{tab:sigma8-w}).

\begin{table}
\caption{\label{tab:sigma8-w} The relation between $w$ and $\sigma_8$.
This relation is derived from Eq.~(\ref{eqn:sigma8-w}).} 
\begin{tabular}{lc}
\hline \hline
$w$ & $\sigma_8$ \\ \hline
$-0.5$ & $0.56967$ \\
$-0.8$ & $0.663688$ \\
$-0.9$ & $0.698358$ \\
$-1.2$ & $0.813615$ \\ \hline
\end{tabular}
\end{table}

In order to avoid the divergence of the density
fluctuation in the limit of large $k$ 
and since the observed distribution is 
intrinsically discrete,
it is necessary to consider the density field 
$\rho(\bm{x};R)$
at the position $\bm{x}$ smoothed over some scale $R$,
which is related to the unsmoothed density field
$\rho(\bm{x})$ as
\begin{eqnarray}
\rho(\bm{x};R) &=& \int d^3 \bm{y} W(|\bm{x}-\bm{y}|;R) 
\rho(\bm{y})
\nonumber\\
&=& \int \frac{d^3 \bm{k}}{(2\pi)^3} \tilde{W} (kR)
\tilde {\rho} (\bm{k}) e^{-i \bm{k}\cdot \bm{x}} \,,
\end{eqnarray}
where $W$ denotes a window function and
$\tilde{W}$ and $\tilde{\rho}$ represent the Fourier transforms
of the corresponding quantities.
In this paper, after the calculations,
we perform smoothing with a spherical top-hat window function,
\begin{eqnarray}
\tilde{W} = \frac{3(\sin x - x \cos x)}{x^3}\,,
\end{eqnarray}
which yield the density fluctuation $\delta(\bm{x};R)$
at the position $\bm{x}$
smoothed over the scale $R$. 
For simplicity we use
$\delta$ to denote $\delta(\bm{x};R)$
unless otherwise stated.

It is also shown that for the case of smoothing with 
the spherical top-hat filter with radius $R$,
in the weakly nonlinear region, the skewness
and kurtosis of the density distribution can be 
obtained analytically
for a given cosmology.
For example, in Einstein-de Sitter Universe,
they are given as
\begin{eqnarray*}
\gamma &=& \frac{34}{7} + 
\frac{d \log \sigma^2 (R)}{d \log R}
+{\cal{O}}(\sigma^2) \, ,
\nonumber\\
\eta &=& \frac{60712}{1323} 
+\frac{62}{3}\frac{d\log [\sigma^2(R)]}{d\log R}
+\frac{7}{3}\left(\frac{d \log [\sigma^2 (R)]}{d \log
	    R}\right)^2\nonumber\\
&&+
\frac{2}{3}\frac{d^2 \log[\sigma^2(R)]}{d (\log R)^2}+
{\cal{O}}(\sigma^2) \, .
\end{eqnarray*}

In our previous paper~\cite{Tatekawa06},
we choose the smoothing scale $R$ as a rather large value
($8 h^{-1}$ Mpc), 
because Lagrangian linear perturbation cannot approximate 
the strongly nonlinear regime well because of 
shell-crossing or caustics formation. 
Since our present analysis is based on  N-body simulations,
we can reproduce small scale structure as well,
which permits us to obtain meaningful results
by choosing smaller values for $R$.

Based on $\delta$ calculated by the procedures above,
we compute the statistical quantities, i.e., 
the dispersion, skewness
and kurtosis
for 6 time slices ($z=5, 4, 3, 2, 1, 0$).
In order to obtain the statistical quantities,
we generate 10 samples for the primordial density
fluctuations by COSMICS for each dark energy model.

It is worth noting that 
we normalized the amplitude of the density spectrum 
at recombination era here, 
while the linear density spectrum 
is normalized by density dispersion 
at present era in other analyses
like \cite{Linder:2003dr,Klypin:2003ug,Kuhlen:2004rw,McDonald:2005gz}.
Therefore in the nonlinear regime, our results seem different
from theirs.
However, as we show later, when $w$ approaches to $0$,
the growth of the fluctuation is suppressed,
which is consistent to them. 

\section{Numerical results} \label{sec:results}

For comparison to the previous results,
we set $R=8 h^{-1} \mbox{Mpc}$
in the comoving Eulerian coordinates 
at the present time ($a=1$). 
In the previous paper, since we applied Lagrangian
linear perturbation, we must examine the degree of 
the validity of the perturbative approach. 

\vspace{3mm}
{\bf (a) Validity of Lagrangian linear perturbation }
\vspace{2mm}

\begin{figure}[tb]
\centerline{
\includegraphics[height=7cm]{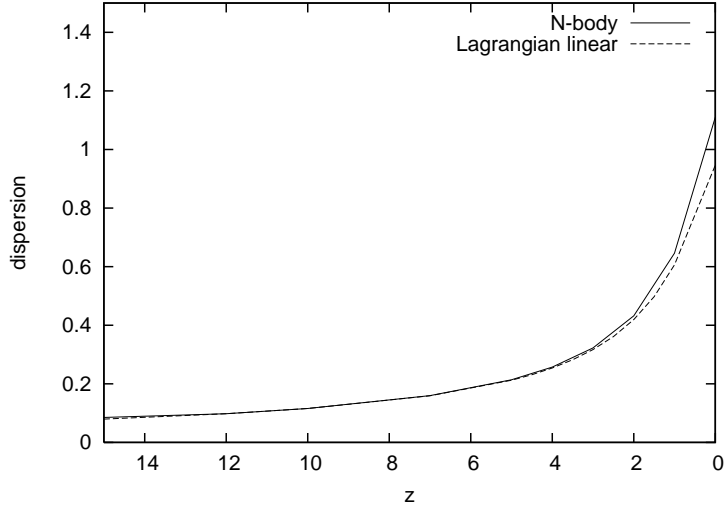}
}
\caption{The dispersion of the density distribution
based on Lagrangian linear perturbation and N-body simulations
($R=8 h^{-1} \mbox{Mpc}$) for $w=-1$.
Lagrangian linear perturbation keeps to be
a good approximation until $z \sim 2$.}
\label{fig:Dispersion}
\end{figure}

At first, we compare the dispersion of the density distribution
obtained by Lagrangian linear perturbation
and N-body simulations. Figure~\ref{fig:Dispersion} shows
the time evolution of the dispersions 
of the density distribution for $w=-1.0$.
In the early stages, since the evolution is well described by
the linear theory, 
the dispersions based on the two methods are almost the same.
For example, the difference is less than $0.7 \%$ at $z=5$. 
In our previous paper, we compare non-Gaussianity of
the density distribution at $z=2$ where the difference 
becomes a little larger, about $3 \%$.

\begin{figure}[tb]
\centerline{
\includegraphics[height=7cm]{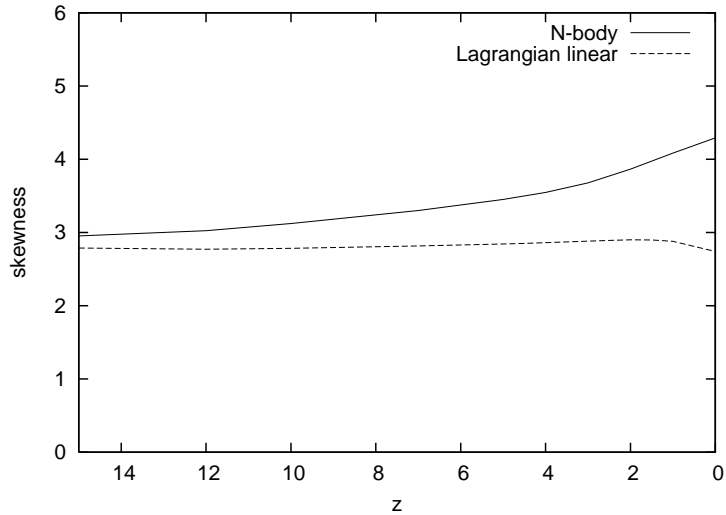}
}
\caption{The skewness of the density distribution
based on Lagrangian linear perturbation and N-body simulations
($R=8 h^{-1} \mbox{Mpc}$) for the case of $w=-1$.
The difference of the skewness which is already
about $5 \%$ at $z=15$ increases.}
\label{fig:Skewness}
\end{figure}

\begin{figure}[tb]
\centerline{
\includegraphics[height=7cm]{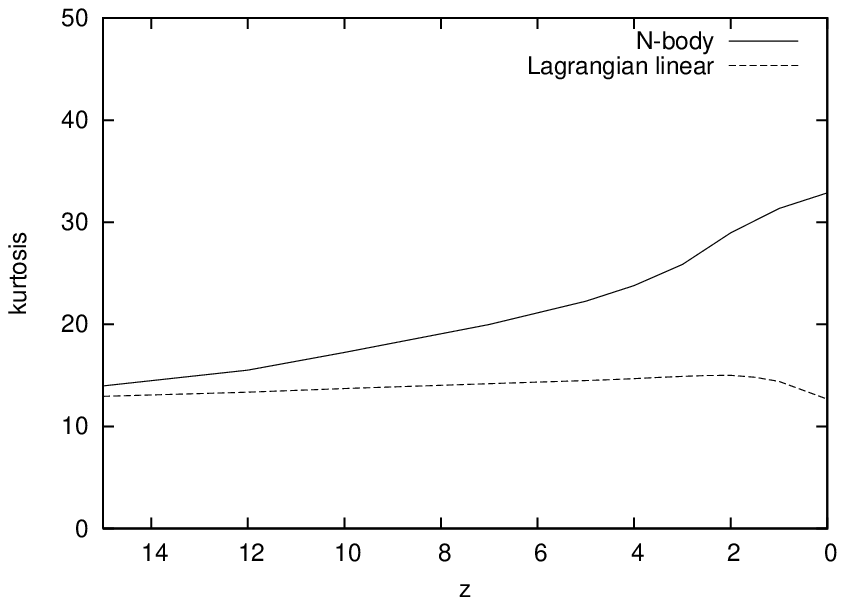}
}
\caption{The Kurtosis of the density distribution
based on Lagrangian linear perturbation and N-body simulations
($R=8 h^{-1} \mbox{Mpc}$) for the case of $w=-1$.
The difference of the kurtosis which is
already about $7 \%$ at $z=15$ increases.}
\label{fig:Kurtosis}
\end{figure}

Even though the difference 
between Lagrangian linear perturbation 
and N-body simulations
is still small until $z=2$ for the dispersion, 
it is not shown to hold also for non-Gaussianity
of the density distribution. 
Figs.~\ref{fig:Skewness} and \ref{fig:Kurtosis}
show the time evolution of the skewness and kurtosis
of the density distribution for  $w=-1$, 
respectively and the difference is about $25 \%$
for the skewness and about $50 \%$ for the kurtosis. 
Therefore, it can be deduced that non-Gaussianity
of the density distribution tend to be underestimated
by Lagrangian linear perturbation, even for the
region where the evolution of the dispersion
is well described by it.

Regardless of the fact above, the conclusion of our 
previous paper that the difference of non-Gaussianity
between $w=-1$ and $w=-0.8$ or $-1.2$
is about $1 \%$ at $z=2$
remains unchanged, since the difference between
Lagrangian linear perturbation and N-body simulations
is also about  $25 \%$ for the skewness and 
about $50 \%$ for the kurtosis for $w=-0.8$ or $-1.2$.
We summarize the difference of the statistical 
quantities between
Lagrangian linear perturbation and N-body simulations
at $z=2$ for various $w$  (Table~\ref{tab:ZA-NBody}).

\begin{table}
\caption{\label{tab:ZA-NBody} The difference of the dispersion,
skewness, and kurtosis between Lagrangian linear
perturbation and N-body simulations at $z=2$ ($R = 8 h^{-1}$ Mpc).} 
\begin{tabular}{lccc}
\hline \hline
$w$ & dispersion & skewness & kurtosis \\ \hline
$-0.5$ & $6.7 \%$ & $-23.4 \%$ & $-45.7 \%$ \\
$-0.8$ & $-1.6 \%$ & $-24.7 \%$ & $-47.8 \%$ \\
$-0.9$ & $-2.5 \%$ & $-24.8 \%$ & $-48.0 \%$ \\
$-1.0$ & $-3.0 \%$ & $-25.0 \%$ & $-48.2 \%$ \\
$-1.2$ & $-3.6 \%$ & $-25.1 \%$ & $-48.4 \%$ \\ \hline
\end{tabular}
\end{table}

\vspace{3mm}
{\bf (b) Statistical quantities ($R = 8 h^{-1}$ Mpc)}
\vspace{2mm}

\begin{figure}[tb]
\centerline{
\includegraphics[height=7cm]{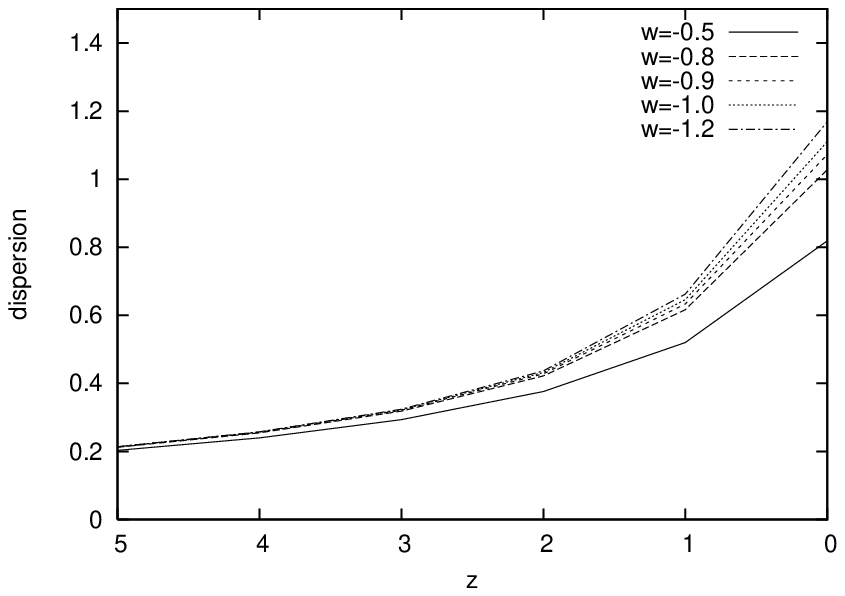}
}
\caption{The dispersion of the density distribution
in several dark energy models ($R=8 h^{-1} \mbox{Mpc}$).
The growth of the dispersion 
continues monotonously. 
As the value of $w$ becomes larger (approaches zero),
the dispersion becomes smaller. 
The difference of the dispersion between $w=-1$ and
$w=-0.8$ is about $2.4 \%$ at $z=2$,
and become as large as  $7.5  \% $ at  $z=0$. 
In the same way, the difference of the dispersion
between  $w=-1$ and $w=-1.2$ is 
 $1.0 \% $ at  $z=2$ and  $5.3 \% $ at  $z=0$.
The difference between
$w=-1$ and $w=-0.5$ is already $13  \% $ at  $z=2$.}
\label{fig:DE-sigmaR8}
\end{figure}

Next we compare the dispersion in several dark energy models
based on N-body simulations. Now that we are using
N-body simulations, we can go into the strongly nonlinear
regime. Figure~\ref{fig:DE-sigmaR8} shows
the time evolution of the dispersion of
the density distribution in several
dark energy models.
The growth of the dispersion monotonously continues
until $z=0$. 
The tendency of the dispersion among dark energy models
is similar to that obtained by
Lagrangian linear perturbation.
As the value of $w$ becomes larger (approaches zero),
the dispersion becomes smaller. This can be explained
as follows: for the model with larger $w$, dark energy
dominates at an earlier era, and the expansion of the
Universe starts accelerating earlier, while it can be 
shown that the growth of the density fluctuation
is slower  in the accelerating stage than in the
matter-dominant stage.

\begin{figure}[tb]
\centerline{
\includegraphics[height=7cm]{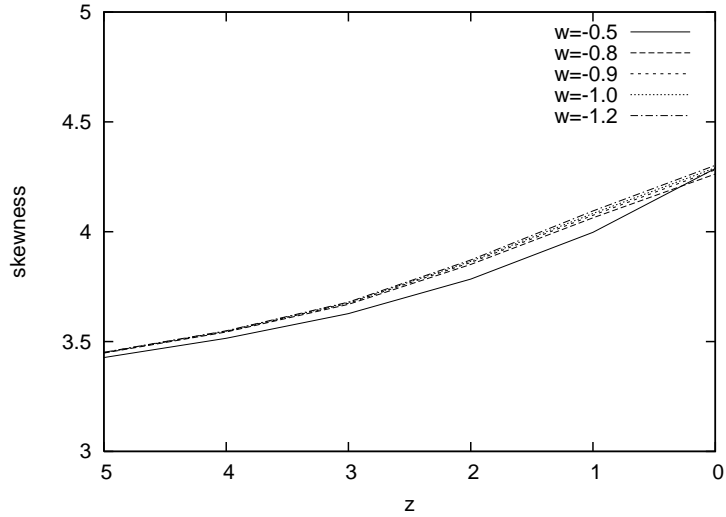}
}
\caption{The skewness of the density distribution
in several dark energy models ($R=8 h^{-1} \mbox{Mpc}$). 
The skewness oscillate between $z=1$ and $z=0$.
We can easily distinguish the case for $w=-0.5$ from
the results.
The difference of the skewness between $w=-1$ and
$w=-0.8$ is about $0.4 \%$ at $z=2$, then spreads out
about $0.7 \%$ at $z=0$. 
In the same way,
the difference of the skewness between $w=-1$ and
$w=-1.2$ is about $0.2 \%$ at $z=2$, then spreads out
about $0.3 \%$ at $z=0$. 
}
\label{fig:DE-skewR8}
\end{figure}

\begin{figure}[tb]
\centerline{
\includegraphics[height=7cm]{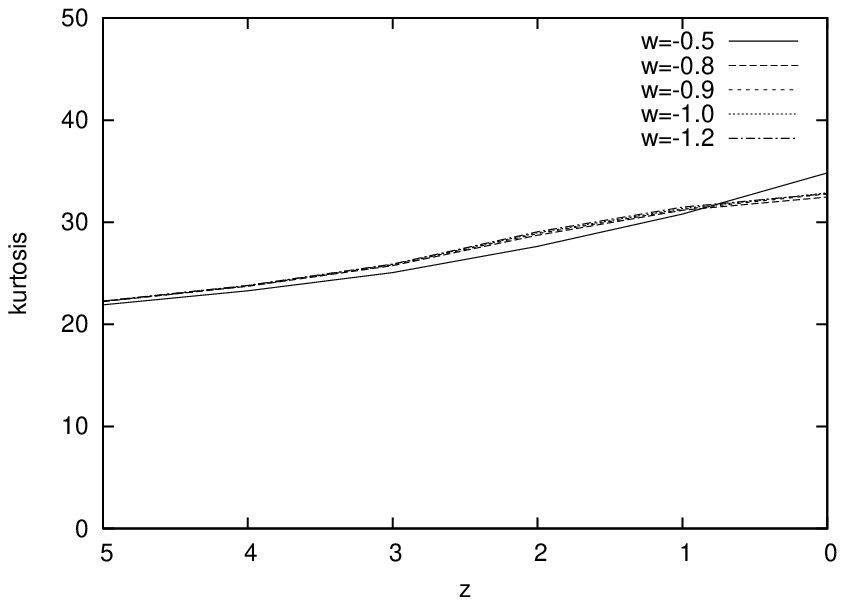}
}
\caption{The kurtosis of the density distribution
in several dark energy models ($R=8 h^{-1} \mbox{Mpc}$).
The kurtosis oscillate between $z=1$ and $z=0$.
As well as the skewness,
the difference of the kurtosis between $w=-1$ and
$w=-0.8$ is about $0.8 \%$ at $z=2$, then spreads out
about $1.2 \%$ at $z=0$. 
In the same way,
the difference of the kurtosis between $w=-1$ and
$w=-1.2$ is about $0.3 \%$ at $z=2$,
then contract to about $0.2 \%$ at $z=1$.
}
\label{fig:DE-kurtR8}
\end{figure}

Figs.~\ref{fig:DE-skewR8} and \ref{fig:DE-kurtR8}
show the time evolution of the skewness and kurtosis
of the density distribution, respectively.
It is different from that of the dispersion, 
in that the growth of them are not monotonous, especially after
$z=1$. 
Even though until $z=1$, like the dispersion, these 
quantities are smaller as the value of $w$ becomes larger,
this order does not hold at the late time.
Except the case $w=-0.5$ whose dispersion is not large enough,
they are larger as the value of $w$ becomes larger.

The detailed difference of the dispersion,
skewness, and kurtosis for $R=8 h^{-1} \mbox{Mpc}$
between $w=-1.0$ and other $w$
at $z=2$ and $z=0$
are summarized in tables ~\ref{tab:z2-R8} and \ref{tab:z0-R8}, 
respectively.

Regardless of the change of the order of the 
$w$ dependence, the differences of both the skewness and
the kurtosis among $w = -1.2$, $w=-1$ and $w=-0.8$ are much
less than $1 \%$ at $z=2$ but turn out to become as large as
$1 \%$ (skewness) and $2 \%$ (kurtosis) at $z=0$. 
For the case of $w=-0.5$, the values of both
the skewness and the kurtosis are obviously less than
that for the case of $w=-1.0$ even at $z=2$
(about $2 \%$ and about $5 \%$, respectively).

\begin{table}
\caption{\label{tab:z2-R8} The difference of the dispersion,
skewness, and kurtosis between $w=-1.0$ and other $w$
at $z=2$ ($R = 8 h^{-1}$ Mpc).} 
\begin{tabular}{lccc}
\hline \hline
$w$ & dispersion & skewness & kurtosis \\ \hline
$-0.5$ & $-13.1 \%$ & $-2.1 \%$ & $-4.6 \%$ \\
$-0.8$ & $-2.4 \%$ & $-0.37 \%$ & $-0.80 \%$ \\
$-0.9$ & $-0.93 \%$ & $-0.14 \%$ & $-0.32 \%$ \\
$-1.2$ & $1.0 \%$ & $0.15 \%$ & $0.34 \%$ \\ \hline
\end{tabular}
\end{table}

\begin{table}
\caption{\label{tab:z0-R8} The difference of the dispersion,
skewness, and kurtosis between $w=-1.0$ and other $w$
at $z=0$ ($R = 8 h^{-1}$ Mpc).} 
\begin{tabular}{lccc}
\hline \hline
$w$ & dispersion & skewness & kurtosis \\ \hline
$-0.5$ & $-26.2 \%$ & $-0.09 \%$ & $6.0 \%$ \\
$-0.8$ & $-7.5 \%$ & $-0.67 \%$ & $-1.2 \%$ \\
$-0.9$ & $-3.4 \%$ & $-0.24 \%$ & $-0.31 \%$ \\
$-1.2$ & $5.3 \%$ & $0.25 \%$ & $-0.19 \%$ \\ \hline
\end{tabular}
\end{table}

\vspace{3mm}
{\bf (c)  Statistical quantities ($R = 2 h^{-1}$ Mpc)}
\vspace{2mm}

In this paper, since we apply N-body simulations, we can
analyze smaller structure. In other words, even if we
choose smaller value for $R$, we can obtain reliable results.
Therefore, next we set the smoothing length 
$R=2 h^{-1} \mbox{Mpc}$
in the comoving Eulerian coordinates 
at the present time ($a=1$). 
Even though the nonlinearity in this scale at $z=0$ is
so strong that the corresponding galaxy distribution 
is influenced by the processes other than
gravity such as the bias related with galaxy formation,
it is worth investigating whether there exist other 
kind of scaling between higher order cumulants and 
the dispersion of the density distribution in this region.

\begin{figure}[tb]
\centerline{
\includegraphics[height=7cm]{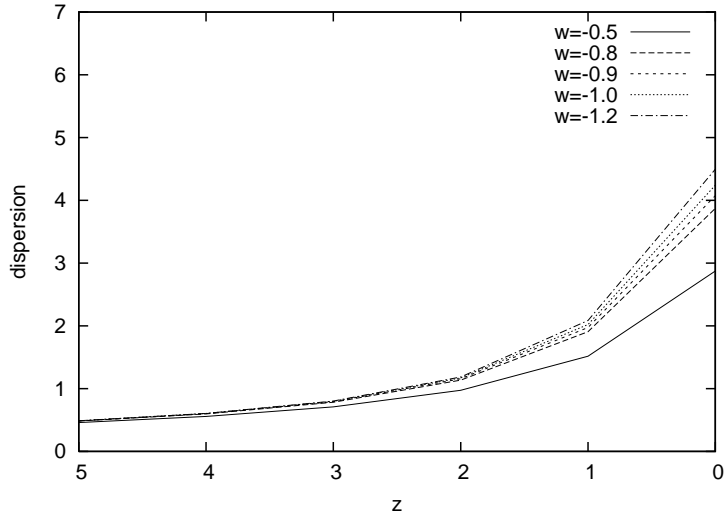}
}
\caption{The dispersion of the density distribution
in several dark energy models ($R=2 h^{-1} \mbox{Mpc}$).
The tendency of the difference of the dispersion
is similar to that in the case of $R=8 h^{-1} \mbox{Mpc}$.
The difference of the dispersion between $w=-1$ and
$w=-0.8$ is about $8.7 \%$ at $z=0$.
Similarly the difference of the dispersion 
between  $w=-1$ and
$w=-1.2$ is about $6.0\%$ at $z=0$.
The difference of the dispersion between $w=-1$ and
$w=-0.5$ is quite large (about $17 \%$ already at $z=2$).
}
\label{fig:DE-sigmaR2}
\end{figure}

Figure~\ref{fig:DE-sigmaR2} shows
the time evolution of the dispersion of
the density distribution in several dark energy models.
The tendency for the growth of the dispersion is similar
to that for $R=8 h^{-1} \mbox{Mpc}$.
The dispersion monotonously grows until $z=0$ and if the
equation of state of dark energy $w$ increases, 
the dispersion would decrease.
Since we choose smaller smoothing scale, 
the dispersion is larger than that in the case of
$R=8 h^{-1} \mbox{Mpc}$ for a fixed $z$ and $w$.

\begin{figure}[tb]
\centerline{
\includegraphics[height=7cm]{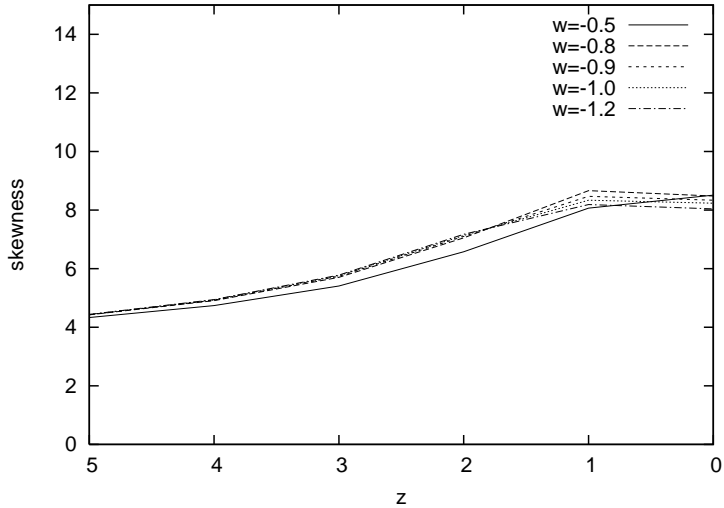}
}
\caption{The skewness of the density distribution
in several dark energy models ($R=2 h^{-1} \mbox{Mpc}$). 
The growth of the skewness stops after $z=1$.
The difference of the skewness between $w=-1$ and
$w=-0.8$ is about $3.0 \%$ at $z=0$.
At some time between $z=1$ and $z=0$,
the skewness of  $w=-0.8$ becomes larger than
that of  $w=-1.0$.
In the same way, the difference between $w=-1$ and
$w=-1.2$ is about $2.4 \%$ at $z=0$.
At some time between $z=1$ and $z=0$,
the skewness of  $w=-1.2$ becomes smaller than
that of  $w=-1.0$.
The difference between $w=-1$ and $w=-0.5$ 
is as large as $3.4 \%$ even at $z=0$. }
\label{fig:DE-skewR2}
\end{figure}

\begin{figure}[tb]
\centerline{
\includegraphics[height=7cm]{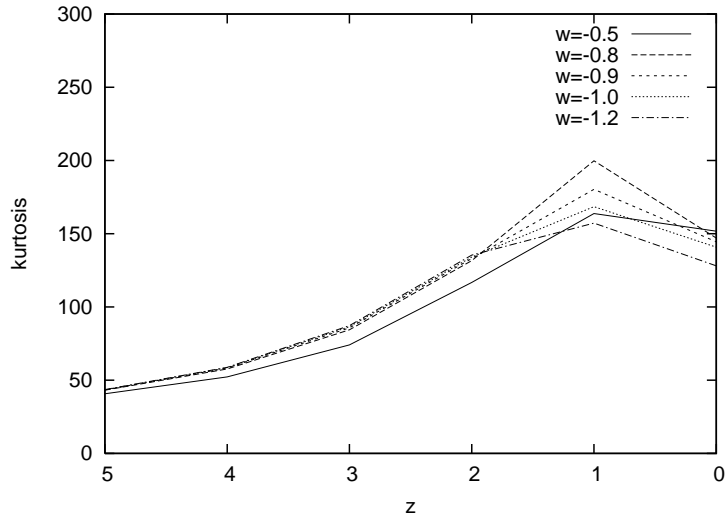}
}
\caption{The kurtosis of the density distribution
in several dark energy models ($R=2 h^{-1} \mbox{Mpc}$).
The growth of the kurtosis stops at $z=1$.
The difference of the kurtosis between $w=-1$ and
$w=-0.8$ is about $4.5 \%$ at $z=0$.
At some time between $z=1$ and $z=0$,
the kurtosis of  $w=-0.8$ becomes larger than
that of  $w=-1.0$.
In the same way, the difference between $w=-1$ and
$w=-1.2$ is about $9 \%$ at $z=0$.
At some time around  $z=2$,
the kurtosis of  $w=-1.2$ becomes smaller than
that of  $w=-1.0$.
The difference between $w=-1$ and $w=-0.5$ 
is as large as $8 \%$ even at $z=0$. 
}
\label{fig:DE-kurtR2}
\end{figure}

Figs.~\ref{fig:DE-skewR2} and \ref{fig:DE-kurtR2}
show the time evolution of the skewness and kurtosis
of the density distribution,
respectively. As in the case of $R=8 h^{-1} \mbox{Mpc}$,
the growth of them is not monotonous.
It is shown that after $z=1$ for a fixed $z$, 
the skewness and kurtosis
increases if the equation of state of dark energy
$w$ increases including the case $w=-0.5$.
This suggests that in sufficiently nonlinear region,
say $\sigma > 2$, there exists some scaling between
the higher order cumulants and the dispersion
$<\delta^n>_c \propto \sigma^{p(n)}$, where 
$p(n) < 2n-2$, at least inside the scope of
our calculations.

The detailed difference of the dispersion,
skewness, and kurtosis 
for $R=2 h^{-1} \mbox{Mpc}$ between $w=-1.0$ and other $w$
at $z=2$ and $z=0$
are summarized in tables ~\ref{tab:z2-R2} and \ref{tab:z0-R2}, 
respectively. 

Regardless of the change of the order of the 
$w$ dependence, the differences of both the skewness and
the kurtosis among $w = -1.2$, $w=-1$ and $w=-0.8$ are much
less than $2 \%$ at $z=2$ but turn out to become as large as
$3 \%$ (skewness) and $9 \%$ (kurtosis)
at $z=0$. 
For the case of $w=-0.5$, the values of both
the skewness and the kurtosis are obviously less than
that for the case of $w=-1.0$ even at $z=2$
(about $8 \%$ and about $13 \%$, respectively).

\begin{table}
\caption{\label{tab:z2-R2} The difference of the dispersion,
skewness, and kurtosis between $w=-1.0$ and other $w$
at $z=2$ ($R = 2 h^{-1}$ Mpc).} 
\begin{tabular}{lccc}
\hline \hline
$w$ & dispersion & skewness & kurtosis \\ \hline
$-0.5$ & $-16.8 \%$ & $-7.8 \%$ & $-13.0 \%$ \\
$-0.8$ & $-3.1 \%$ & $-1.2 \%$ & $-1.9 \%$ \\
$-0.9$ & $-1.2 \%$ & $-0.48 \%$ & $-0.81 \%$ \\
$-1.2$ & $1.3 \%$ & $0.53 \%$ & $0.97 \%$ \\ \hline
\end{tabular}
\end{table}

\begin{table}
\caption{\label{tab:z0-R2} The difference of the dispersion,
skewness, and kurtosis between $w=-1.0$ and other $w$
at $z=0$ ($R = 2 h^{-1}$ Mpc).} 
\begin{tabular}{lccc}
\hline \hline
$w$ & dispersion & skewness & kurtosis \\ \hline
$-0.5$ & $-32.2 \%$ & $3.4 \%$ & $7.7 \%$ \\
$-0.8$ & $-8.7 \%$ & $3.0 \%$ & $4.5 \%$ \\
$-0.9$ & $-3.9 \%$ & $1.2 \%$ & $2.6 \%$ \\
$-1.2$ & $6.0 \%$ & $-2.4 \%$ & $-9.0 \%$ \\ \hline
\end{tabular}
\end{table}

\vspace{3mm}
{\bf (d) Probability distribution function}
\vspace{2mm}

\begin{figure}[tb]
\centerline{
\includegraphics[height=7cm]{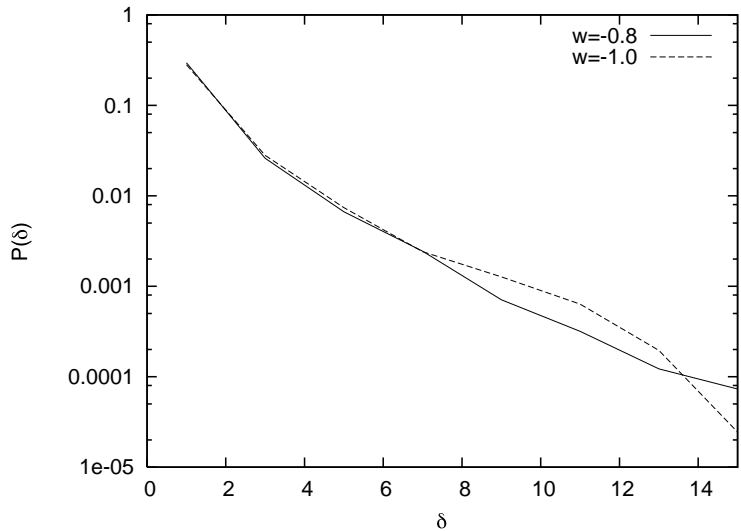}
}
\caption{The PDF of the density distribution
in $w=-0.8$ and $w=-1.0$ at $z=0$ ($R = 8 h^{-1}$ Mpc).
Because of smoothing for the PDF, we pick up the PDF
with low resolution ($\Delta \delta=2$).
Since the PDF of $w=-1.0$ spreads more widely than that of
$w=-0.8$ for almost the same dispersion,
the non-Gaussianity of the PDF of $w=-1.0$ is larger
than that of $w=-0.8$.
}
\label{fig:P-delta-R8}
\end{figure}

\begin{figure}[tb]
\centerline{
\includegraphics[height=7cm]{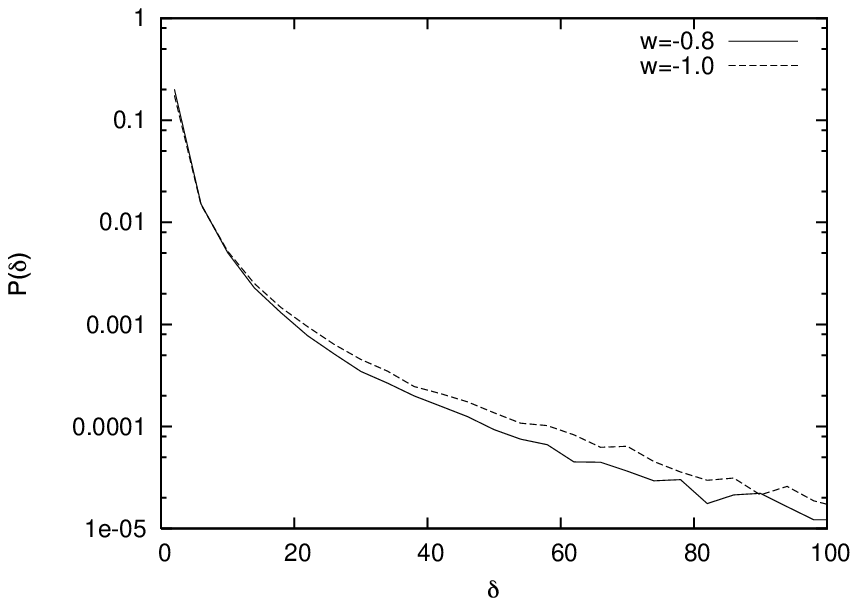}
}
\caption{The PDF of the density distribution
in $w=-0.8$ and $w=-1.0$ at $z=0$ ($R = 2 h^{-1}$ Mpc).
Because of smoothing for the PDF, we pick up the PDF
with low resolution ($\Delta \delta=4$).
Even though the PDF of $w=-1.0$ spreads more widely than
that of $w=-0.8$, since the dispersion of the PDF of $w=-1.0$
is larger than that of $w=-0.8$, the resulting non-Gaussianity
becomes smaller for the case of $w=-1.0$.
}
\label{fig:P-delta-R2}
\end{figure}

In order to clarify the difference between the case of
$w=-0.8$ and $w=-1.0$, 
we present raw data of PDF of the density distribution
at $z=0$
for $R=8h^{-1}$Mpc (Figure~\ref{fig:P-delta-R8}),
and $R=2h^{-1}$Mpc (Figure~\ref{fig:P-delta-R2}),
respectively. 
When the nonlinear effect promotes the formation of the dense
structure and restrains the evolution of the voids,
the degree of asymmetry of PDF (skewness)
grows obviously. 
On the other hand, 
if the edge of PDF spreads widely without reduction of peak,
the ``peakiness'' of PDF (kurtosis) grows evidently.

\vspace{3mm}
{\bf (e) Error analysis}
\vspace{2mm}

\begin{figure}[tb]
\centerline{
\includegraphics[height=7cm]{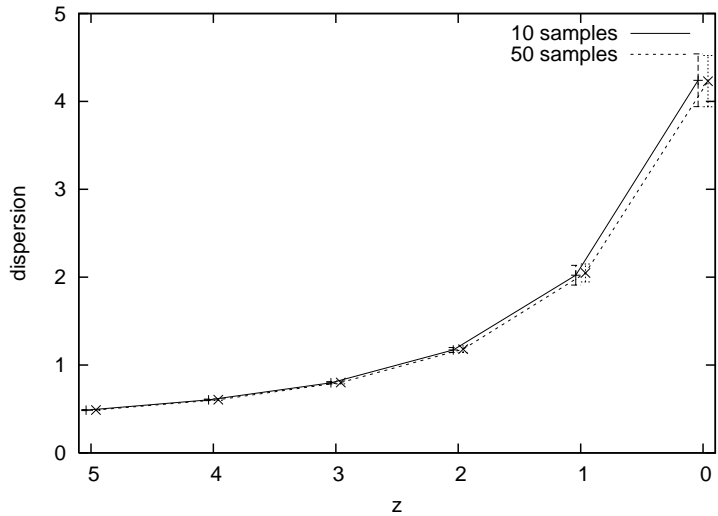}
}
\caption{The dispersion of the density distribution
in the case of $w=-1.0$ ($R=2 h^{-1} \mbox{Mpc}$).
Here we analyze the variance of samples.
When we increase number of samples, the variance slightly
decreases. 
The variance is about $7 \%$ at $z=0$ in both cases.
}
\label{fig:w10-Sigma-N50}
\end{figure}

\begin{figure}[tb]
\centerline{
\includegraphics[height=7cm]{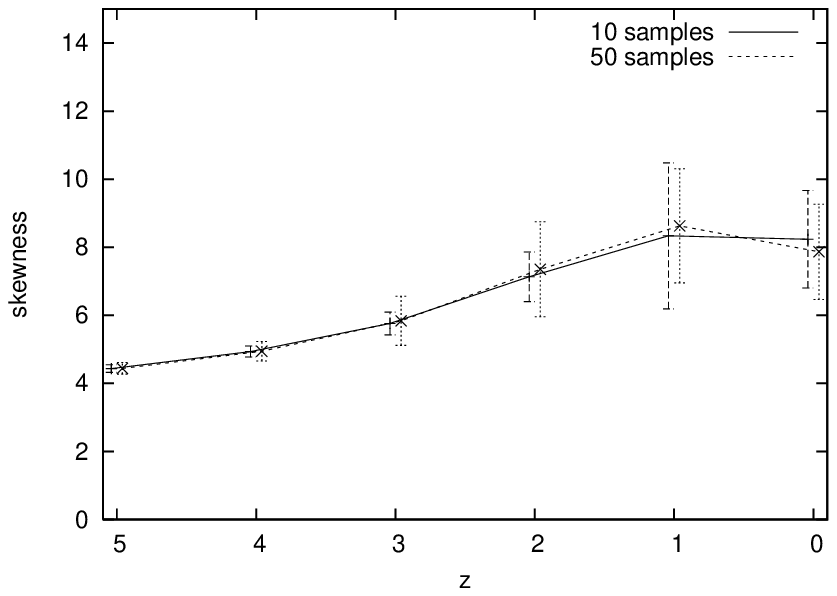}
}
\caption{The skewness of the density distribution
in the case of $w=-1.0$ ($R=2 h^{-1} \mbox{Mpc}$).
Here we analyze the variance of samples.
Even if we increase number of samples, the variance of the
non-Gaussianity is not always improved.
The variance is about $18 \%$ at $z=0$.
}
\label{fig:w10-Skew-N50}
\end{figure}

\begin{figure}[tb]
\centerline{
\includegraphics[height=7cm]{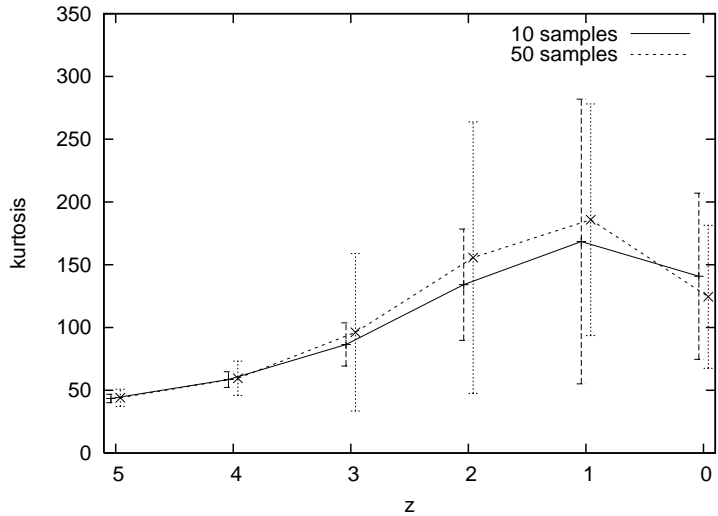}
}
\caption{The kurtosis of the density distribution
in the case of $w=-1.0$ ($R=2 h^{-1} \mbox{Mpc}$).
Here we analyze the variance of samples.
Even if we increase number of samples, the variance of the
non-Gaussianity is not always improved.
The variance is about $46 \%$ at $z=0$.
}
\label{fig:w10-Kurt-N50}
\end{figure}

Until now, we provided our results without error bars, 
even though they are obtained by averaging 10
samples for each dark energy model.
However, in order to check whether these results
can be used directly as a cosmological tool,
the statistical error analysis is necessary.
Here we estimate the variances among samples
for the case of $w=-1.0$
by increasing the number of samples up to 50.
The results are showed
in Figs.~\ref{fig:w10-Sigma-N50}-\ref{fig:w10-Kurt-N50}.
As a result, contrary to our expectation, the variance
is much larger than the difference between the difference
of the non-Gaussianity between $w=-1$ and the other cases.
Furthermore, it turns out that 
although the variance of the dispersion 
can be decreased , the variance of the
non-Gaussianity is not necessarily  improved
by increasing the number of samples. 
The detailed results are shown in Figs.~\ref{fig:w10-Sigma-N50}-\ref{fig:w10-Kurt-N50}.
Therefore, for the practical use of this analysis
to constrain the dark energy models, further improvements
for the calculation of the statistical quantities,
which seems to be worth considering is necessary.

\section{Summary} \label{sec:summary}

To clarify the nature of dark energy is, without doubt,
one of the most important tasks in modern cosmology.
From the phenomenological viewpoint, 
it is important to constrain the effective equation of state
of dark energy $w$ by observations. Even though we
have obtained the constraint as $w < -0.90$ 
(95 \% confidence limit assuming $w \ge -1$)~\cite{WMAP} 
by combining WMAP data
with other astronomical data, to pin down 
the value of $w$,
it is necessary to propose
other independent observational tools.

For this purpose, we consider non-Gaussianity of
the density distribution which is generated 
by the non-linear dynamics. In this paper, as a natural
extension to our previous work \cite{Tatekawa06} where
we follow a semi-analytic approach based on Lagrangian
linear perturbation theory, we apply N-body simulations
based on $P^3 M$ codes. In terms of the density
fluctuation obtained by N-body simulations, we compute
the skewness and kurtosis of the smoothed density
distribution with the appropriate definition in that
they become constant in extremely weakly nonlinear stage.
They are important statistical quantities denoting the
deviation from the Gaussian distribution.
By considering several constant-$w$
dark energy models ($w=-0.5, -0.8, -0.9, -1.0, -1.2$),
we present the $w$-dependence of the time evolution
of non-Gaussianity of the density distribution.

First we compare the results for the weakly nonlinear region
at $z=2$ obtained by N-body simulations
and Lagrangian linear perturbation
by choosing the same smoothing length
$R = 8 h^{-1}$ Mpc.
In this region where non-Gaussianity
become smaller as $w$ becomes larger, 
we regarded the results by Lagrangian
linear perturbation as reliable in the previous work. 
As a result, even though the dispersion obtained 
by Lagrangian linear
perturbation agrees well with the ones obtained by
N-body simulations until $z\sim2$, non-Gaussianity
tend to be underestimated by Lagrangian linear perturbation.

This result is attributed to the fact that 
second order Lagrangian perturbation is necessary
to express second order Eulerian perturbation
\cite{Grinstein87,Bouchet92}, mathematically
and the peculiar acceleration perpendicular to
the peculiar velocity can not be picked up by 
Lagrangian linear perturbation, physically.
Therefore, we learn the fact 
that ``Lagrangian linear perturbation
is not necessarily better than second order Eulerian
perturbation''.
However, regardless of the underestimation of 
non-Gaussianity, the conclusions of the previous paper
that the relative difference of non-Gaussianity between 
$w=-1.0$ and $w=-0.8$ or $w=-1.2$ is about  $1 \%$
at $z=2$ for $R=8 h^{-1} \mbox{Mpc}$ remains unchanged
because the degree of the underestimation in Lagrangian linear
perturbation is almost $w$-independent.

Next, we analyze the statistical quantities in 
the strongly nonlinear region where the previous
perturbative approach is obviously invalid.
We find that the relative difference of non-Gaussianity
among several constant-$w$ models spread as $z\to 0$,
even though the growth of non-Gaussianity is not monotonous
in that region and the order changes around $z=0.5$
except $w=-0.5$ model whose dispersion is not so large enough.
For example, at $z=0$ the relative difference between 
non-Gaussianity of $w=-0.8$ model and 
that of $w=-1.0$ model is $0.67 \%$ for the skewness
and $1.2 \%$ for the kurtosis,
while at $z=2$ they are $-0.52 \%$ for the skewness
and  $-1.3 \%$ for the kurtosis.
Similarly, at $z=0$ the difference 
between non-Gaussianity
of $w=-1.2$ model and that of $w=-1.0$ model is 
$0.25 \%$ for the skewness and $-0.19 \%$ for the kurtosis,
while at $z=2$ they are $0.15 \%$ for the skewness
and $0.34 \%$ for the kurtosis.

Making use of N-body simulations, we also analyze
the smaller structure by setting the smoothing scale
$R=2 h^{-1} \mbox{Mpc}$, even though the nonlinearity
in this scale at $z=0$ is so strong that 
the galaxy distribution
is influenced by the process other than gravity. 
For this case, since we look at
the smaller scale, the dispersion and non-Gaussianity
of the density distribution is larger than that in the case
of $R=8 h^{-1} \mbox{Mpc}$ for a fixed $z$ and $w$.
Relative difference of non-Gaussianity among 
several constant-$w$ models are also greater than 
the case of $R=8 h^{-1} \mbox{Mpc}$.

For example, at $z=0$ the relative difference between  
non-Gaussianity of $w=-0.8$ model and 
that of $w=-1.0$ model is $3.0 \%$ for the skewness
and $4.5 \%$ for the kurtosis. Similarly, 
the relative difference between the 
non-Gaussianity of $w=-1.2$ model and 
that of $w=-1.0$ model is $-2.4 \%$ for the skewness
and $-9.0 \%$ for the kurtosis.
Furthermore, on the contrary to the weakly nonlinear region,
for a fixed $z$ we confirm non-Gaussianity become larger 
as $w$ becomes larger after $z=0.5$, 
including the case of $w=-0.5$.

This suggests that in the sufficiently nonlinear region,
say $\sigma > 2$, there exists some scaling between
the higher order cumulants and the dispersion
$<\delta^n>_c \propto \sigma^{p(n)}$, where 
$p(n) < 2n-2$, at least inside the scope of
our calculations. Academically, not limited to 
the constraint for the dark energy models,
this result seems to worth investigating related with 
the nature of the nonlinear gravitational
clustering in the expanding Universe.

It is necessary to mention the possibility of examining
our results. 
Recently, several galaxy redshift survey projects have been
progressing~\cite{Hawkins03,Abazajian04,Gerke04}. 
In these projects,
many galaxies within a region ($z<0.3$ for 2dF,
$z<0.5$ for SDSS, and $0.7 < z < 1.4$ for DEEP2) have
been observed. These projects show the latest distribution of
galaxies, which form strongly nonlinear structure.
Actually, the skewness and kurtosis of the distribution
of SDSS Early Data Release had been computed 
\cite{Szapudi:2001mh}. The results suggested the SDSS
imaging data can enable to determine the skewness
and kurtosis up to $1 \%$ and less than $10 \%$,
respectively. Therefore, in future projects we can
expect to determine the skewness and kurtosis with high 
accuracy. However, based on our error analysis, 
the variance of the non-Gaussianity turned out to be
about $50 \%$ for 10 samples and it does not increased
even by increasing the number of samples up to 50.
This suggests that our analysis, by itself,
is still immature and further improvements are necessary 
for the practical use to constrain the property of 
dark energy.

As mentioned above, for smaller scale like 
$R=2 h^{-1} \mbox{Mpc}$ now, the complex physical processes
about galaxy formation affect the galaxy distribution.
In order to use observational data such as the PSCz
survey \cite{Saunders:2000af} which is an all sky survey at the present epoch,
it is necessary to create mock PSCz catalogs from our results
making use of the PSCz selection function 
\cite{Rowan-Robinson:1999mx}.
Even though this is from excessively academical interest,
$w$-dependence of the non-Gaussianity of the actual galaxy
distribution seems to be worth investigating.
On the other hand, it is known that weak lensing
surveys can potentially provide us with precision maps
of the projected density up to redshifts around $1$
\cite{vanWaerbeke:2000rm,Bacon:2000sy,Wittman:2000tc,Kaiser:2000if}. 
Even though we need another step of obtaining
the convergence field which can be written as the 
projection of the matter density along the line of sight,
the skewness and kurtosis of the convergence field
can be tested by weak lensing surveys.
Obviously, our results obtained in this paper help
to derive these quantities.

Finally, it is worth noting that primordial non-Gaussianity
which is generated by inflation \cite{BKMR}. 
From WMAP observations,
a limit for the non-linear coupling parameter has been 
established \cite{Komatsu:2003fd} which does not
deny the existence of non-Gaussianity in the primordial
density fluctuations. Even though we cannot disentangle
them generally, we can calculate non-Gaussianity
of large scale structure for a given non-Gaussian initial
condition and compare the results with the one obtained
by the Gaussian initial condition. This requires
further investigations and we hope to report results
in a separate publication.

\ack
We thank to Masahiro Morikawa for useful discussion.
For usage of COSMICS and $P^3M$ codes, we would like to thank
Edmund Bertschinger and Alexander Shirokov.
SM is supported in part by the Japan Society for Promotion
of Science (JSPS) Research Fellowship.


\begin{thebibliography}{99}
\bibitem{PNG}
However, it must be worth noting that 
the small deviation from Gaussian
pertrubation at primordial time 
become one of the main topic
for the research of Early Universe,
see for example \cite{BKMR}
\end{thebibliography}
\end{document}